 \documentclass[final,5p,times,twocolumn]{elsarticle}

\usepackage{amssymb}
\usepackage{amsthm}

\usepackage{lineno}
\usepackage{multirow}

\journal{Physics Letters B}

\begin{document}

\begin{frontmatter}



\title{Nuclear force and the EMC effect}

\author{Rong Wang$^{a,b,c}$, Xurong Chen$^{a}$}

\address{
$^a$ Institute of Modern Physics, Chinese Academy of Sciences, Lanzhou 730000, China \\
$^b$ Lanzhou University, Lanzhou 730000, China \\
$^c$ University of Chinese Academy of Sciences, Beijing 100049, China \\
}

\begin{abstract}
A linear correlation is shown quantitatively between the magnitude of the EMC effect
measured in electron deep inelastic scattering (DIS) and the nuclear residual strong
interaction energy (RSIE) obtained from nuclear binding energy subtracting the
Coulomb energy contribution. This phenomenological relationship is used to extract
the size of in-medium correction (IMC) effect on deuteron and to predict the EMC
slopes $|dR_{EMC}/dx|$ of various nuclei. We further investigate the correlations
between RSIE and other quantities which are related to the EMC effect. The observed
correlations among RSIE, EMC slope and SRC ratio $R_{2N}N_{total}/N_{np(^3S_1)}$ imply
that the local nuclear environment drives the modification of quark distributions.
\end{abstract}

\begin{keyword}
Nuclear force \sep EMC effect
\PACS 25.30.Fj \sep 13.60.Hb \sep 21.10.Dr \sep 21.30.-x
\end{keyword}

\end{frontmatter}

%%%\linenumbers

\section{Introduction}

The per-nucleon structure function $F_2^A$ measured on a nucleus ($A>2$) was
first reported to be smaller than that measured on deuterium at intermediate
$x_B$ ($0.35< x_B< 0.7$) by European Muon Collaboration in 1983 \cite{emc-discovery}.
This phenomenon is now commonly referred to as the EMC effect, which was completely
unexpected before the experiment. The early expectation was that the per-nucleon
lepton deep inelastic scattering (DIS) cross sections of heavy nuclei would not
differ much from that of deuteron, for the nuclear binding energies are subtle
compared to the high energy lepton probes (at GeV energy scale and higher).
Anyway, the quark momentum distributions in bound nucleons embedded in nuclei
are modified. A lot of theoretical efforts have been made aimed at understanding
the underlying physics which alters the quark distributions inside nuclei.
Comprehensive reviews of the EMC effect can be found in Refs.
\cite{review1,review2,review3,review4}. However, there is no generally accepted
model for the effect over all $A$ and $x_B$.

Early experiment at SLAC showed that the EMC effect was logarithmically related
to atomic mass number $A$, or proportional to the average nuclear density \cite{emc-slac}.
However, recent measurement at JLab found that the assumption that the size of the EMC
effect scales with nuclear density breaks down for very light nuclei \cite{emc-jlab}.
It is suggested that the effect scales with the local nuclear environment of the
nucleons. Since then, the nuclear dependence of quark distributions has become
an important subject to explore the origin of the EMC effect.

Detailed analysis of the nuclear dependence of the EMC effect and short-range correlations
(SRC) is presented in Ref. \cite{detail-emc-src}, aimed at testing the possible explanations
for the correlation between the EMC effect and SRC \cite{src-emc1}. It is suggested that
the local density explanation \cite{emc-jlab,detail-emc-src} is slightly better than the explanation
in terms of high virtuality \cite{src-emc1,src-emc2} by comparing the fits to EMC slopes
versus $a_2$ and the fits to EMC slopes versus $R_{2N}N_{total}/N_{iso}$.
The SRC scaling factor $a_2=(2/A)\sigma_A/\sigma_d$ is defined as the ratio of per-nucleon
inclusive electron scattering cross section on nucleus A to that on deuteron
at $Q^2>1.4$ (GeV$^2/c^2$) and $1.5<x_B<1.9$ \cite{jlab-src}. $R_{2N}$ is similar to $a_2$
but with the correction for c.m. motion of the correlated pair, which better represents
the relative probability of a nucleon being part of a short-range correlation pair \cite{detail-emc-src}.
The explanation for the EMC-SRC correlation is still not clear.

As we know, protons and neutrons inside nuclei are bound together with nuclear force.
In Quantum Chromodynamics (QCD) theory level, the powerful attractive nuclear force
comes from the residual strong interaction of quarks, which resembles the Van der
Waals force between molecules. The emergence of nuclear force from QCD theory is a
complex phenomenon and depends on the distances being considered. Nonetheless,
calculations of interactions among nucleons are starting to be realized from Lattice QCD
\cite{LQCD1,LQCD2,LQCD3}. The nuclear medium modifies the quark distributions of a nucleon.
A related fundamental question is wether the nuclear force plays an important role
in the EMC effect.

Role of nuclear binding in the EMC effect is an important and old issue
\cite{G.L.Li,F.Gross}. Usually, the contribution of nuclear binding
is believed to be small in the convolution picture \cite{G.L.Li}.
The EMC effect can not be explained by nuclear binding and nucleon Fermi motion alone.
Nevertheless, with new correction to the convolution formula, some \cite{F.Gross}
argue that the nuclear binding effects may be sufficient to explain the EMC effect
at intermediate $x_B$. Recent phenomenological study of the nuclear structure function
perfectly describes the EMC slopes and detailed shapes of the EMC effect
with a few parameters \cite{Kulagin06}. In the model, a number of different nuclear effects
including nuclear shadowing, Fermi motion and binding, nuclear pion excess and
off-shell correction to bound nucleon structure functions are considered.
The off-shell correction is quantified by the average nucleon separation energy.
Interestingly, model calculations including off-shell effect are in agreement with
the structure functions for light nuclei \cite{Kulagin10,Omar}.

Instead of studying the contribution of nuclear binding to the EMC effect,
we try to find out the relation between nuclear force and the EMC effect.
In this work, we define nuclear residual strong interaction energy (RSIE)
as the energy (mass) loss of nucleons binding together with nucleon-nucleon
strong interaction (see Sec. \ref{sec2}). The linear correlation between RSIE and
the EMC effect is shown in Sec. \ref{sec3}. In Sec. \ref{sec4}, we investigate the
correlations between RSIE and other quantities which are connected to the EMC effect.
Finally, a summary is given in Sec. \ref{sec5}.

\section{Nuclear binding and residual strong interaction energy}
\label{sec2}

The simplest description of strength of nuclear force is the nuclear binding energy.
The nuclear binding energy, defined as $B=ZM(^1H)+NM(^1n)-M(A,Z)$, is one of
the well known static properties of the nuclei. In experiment, the nuclear binding
energy is precisely measured for most of the nuclei \cite{atomic-mass}.
While it is known in theory that nuclear binding results from the dominated strong
interaction, it is difficult to calculate. A simple estimation of the nuclear binding energy
can be calculated by the semi-empirical Bethe-Weizs\"{a}cker (BW) mass formula \cite{BW1,BW2}.
According to BW formula, the binding energy of a nucleus of atomic mass number
$A$ and proton number $Z$ is described as
\begin{equation}
B(A,Z)=a_vA-a_sA^{2/3}-a_cZ(Z-1)A^{-1/3}-a_{sym}(A-2Z)^2A^{-1}+\delta,
\label{BWMF}
\end{equation}
where $a_v=15.79$ MeV, $a_s=18.34$ MeV, $a_c=0.71$ MeV, and $a_{sym}=23.21$ MeV.
The pairing energy $\delta=+a_pA^{-1/2}$ for even $N$-even $Z$, $-a_pA^{-1/2}$
for odd $N$-odd $Z$, and $0$ for odd A nuclei, with $a_p=12$ MeV.

Similar to nuclear binding energy, we define RSIE (abbreviation of residual strong
interaction energy) as the energy loss due to the nucleon-nucleon strong interaction.
The definition is given to quantitatively describe the strength of strong interaction
part between nucleons. Assuming the nuclear binding comes from only electro-magnetic
and strong interaction, RSIE can be extracted from binding energy after
Coulomb contribution $-a_cZ(Z-1)A^{-1/3}$ removed. Hence, we get
\begin{equation}
RSIE(A,Z)=B(A,Z)+a_cZ(Z-1)A^{-1/3}.
\label{RSIE-formula}
\end{equation}
The charged protons inside a nucleus repel each other resulting in Coulomb repulsive energy.
The Coulomb energy is evaluated by taken the nucleus as a liquid spherical
charged drop and with self-Coulomb energy of $Z$ protons removed, which is a good
approximation for heavy nuclei. The very light nuclei may not exist as a spherical
charged drop, yet $a_cZ(Z-1)A^{-1/3}$ is still a simple estimation. 
Although BW mass formula is of excellent accuracy for heavy nuclei,
it fails in describing very light nuclei and nuclei with magic number.
Therefore we take the measured binding energies $B(A,Z)$ from experiments in this analysis
instead of the calculations from BW formula. Besides, the measured binding
energies are precise and model independent.

The obtained per-nucleon RSIE of various nuclei are shown in Table \ref{table1}
by using Eq. \ref{RSIE-formula} and nuclear binding energy data from experimental measurements
\cite{atomic-mass}. For a chemical element, the binding energy varies with the isotope.
In the EMC effect experiments, some of the target elements have more than one stable isotope.
For the cases of Cu and Ag, two stable isotopes both have big natural abundances (see Table \ref{table1}).
The natural abundances are taken from Ref. \cite{nndc}. Although the binding energy differences
are small for the isotope, the mean binding energy and the mean mass number of Cu and Ag are used.

\begin{table}[htp]
\begin{center}
\caption{
Some of the measured or calculated quantities of the studied nuclei. Column 2, 3 and 4
show the data of binding energy per nucleon \cite{atomic-mass},
RSIE per nucleon, and the natural abundance \cite{nndc}, respectively.
}
\begin{tabular}{cccc}
\hline
\multirow{2}*{Nucleus} & Binding energy & RSIE / A   & Natural  \\
                       & / A (MeV)      & (MeV) & abundance    \\
\hline
Deuteron   & 1.112 & 1.112 & -     \\
$^3$He     & 2.573 & 2.901 & -   \\
$^4$He     & 7.074 & 7.298 & 99.999866\%  \\
$^9$Be     & 6.463 & 6.918 & 100\%        \\
$^{12}$C   & 7.680 & 8.455 & 98.93\%      \\
$^{27}$Al  & 8.332 & 9.699 & 100\%       \\
$^{40}$Ca  & 8.551 & 10.52 & 96.94\%      \\
$^{56}$Fe  & 8.790 & 10.94 & 91.754\%     \\
$^{63}$Cu  & 8.752 & 11.05 & 69.15\%     \\
$^{65}$Cu  & 8.757 & 10.96 & 30.85\%     \\
$^{107}$Ag & 8.554 & 11.58 & 51.839\%     \\
$^{109}$Ag & 8.548 & 11.50 & 48.161\%     \\
$^{197}$Au & 7.916 & 11.73 & 100\%        \\
\hline
\end{tabular}
\label{table1}
\end{center}
\end{table}

\section{Nuclear force and the EMC effect}
\label{sec3}

The strength of the EMC effect is taken as $|dR_{EMC}/dx|$ \cite{emc-jlab},
which is the slope value of the cross section ratio $R_{EMC}$ from a linear fit
in the intermediate $x_B$ range from 0.35 to 0.7. This definition of the magnitude
of the EMC effect is largely unaffected by the normalization uncertainties,
which is better than taking the ratio $R_{EMC}$ at a fixed value of $x_B$.
Combined data \cite{detail-emc-src} of the measured EMC slopes in electron DIS
at SLAC \cite{emc-slac} and Jlab \cite{emc-jlab} are determined by J. Arrington et al.
Recently, S. Malace et al. \cite{review4} extracted the measured EMC slopes of
various nuclei from global fits to many more experimental data including earlier
measurements by HERMES, NMC, EMC, BCDMS and Rochester-SLAC-MIT collaborations.
EMC slopes from both J. Arrington et al. and S. Malace et al. are taken in this analysis.

\begin{figure}[htp]
\centering
\includegraphics[width=0.46\textwidth]{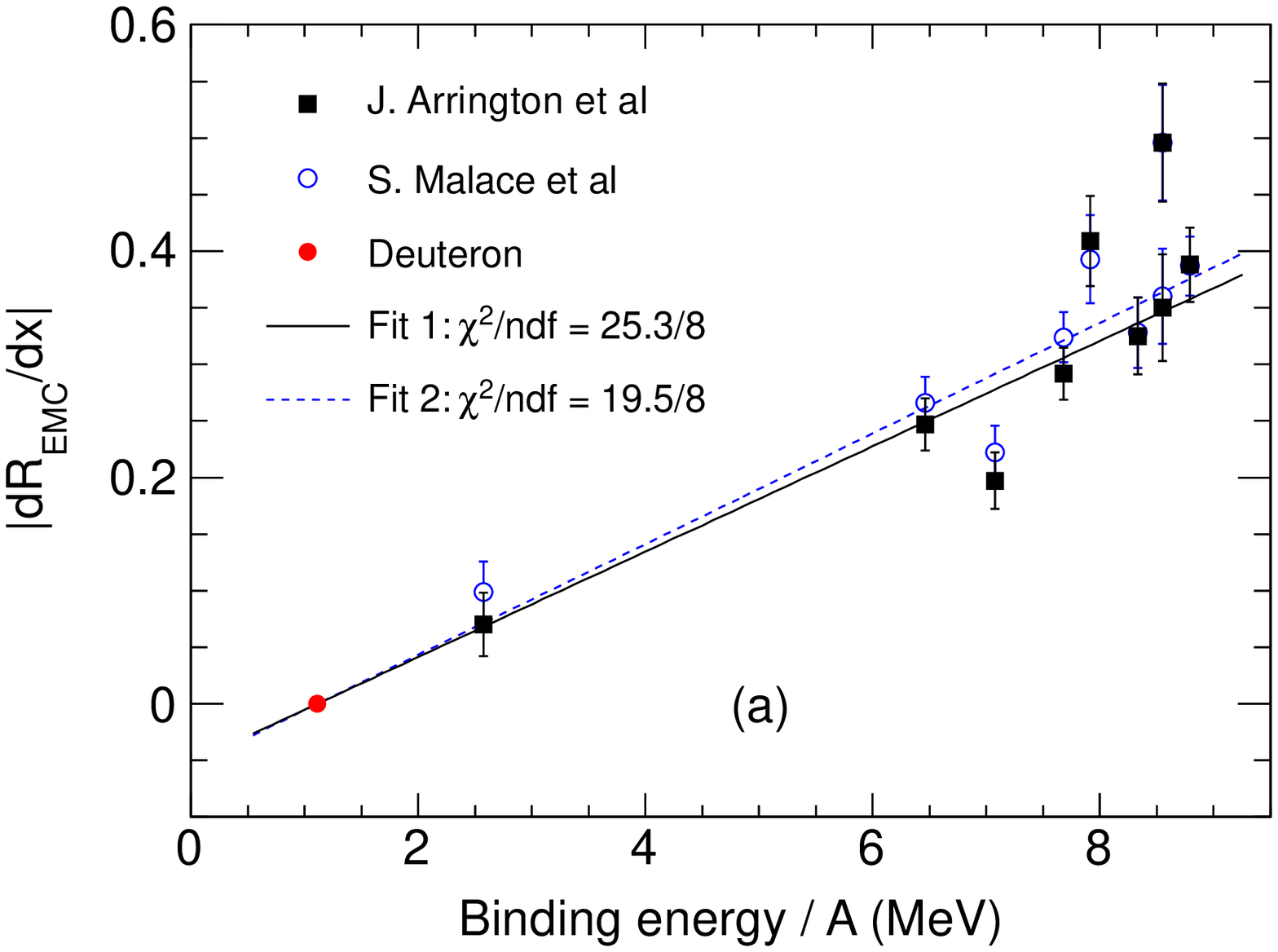}
\includegraphics[width=0.46\textwidth]{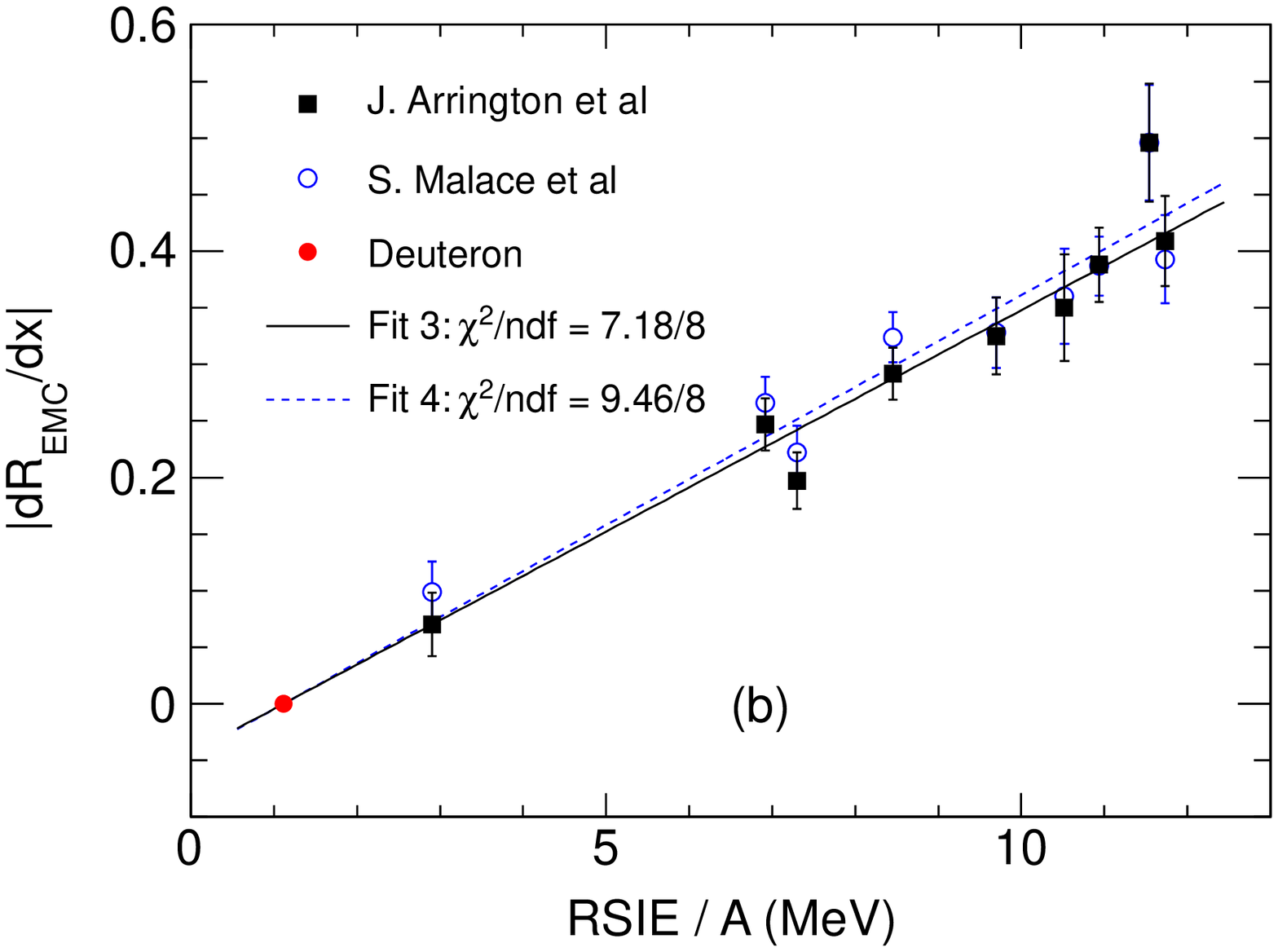}
\caption{
(Color online.)
Plot (a) shows the EMC slopes versus the per-nucleon nuclear binding energies,
and plot (b) shows the EMC slopes versus the per-nucleon RSIEs.
The black solid lines (Fit 1 and 3) are the linear fits to the data from J. Arrington et al.
with a constraint by the deuteron data. The blue dashed lines (Fit 2 and 4) are the linear
fits to the data from S. Malace et al. with a constraint by the deuteron data.
}
\label{fig1}
\end{figure}

The EMC slopes of different nuclei are shown in Fig. \ref{fig1}(a) as a function
of nuclear binding energy per nucleon. The correlation between this two quantities
is not obvious. Linear fits to the data with the constraint by the deuteron (red dot)
are shown in the figure. There is only one free parameter in the fits which is the slope
of the linear function. Fig. \ref{fig1}(b) shows the EMC slopes versus per-nucleon RSIEs.
Strikingly, a clear linear correlation shows up between these two quantities.
The solid line and dashed line in the plot are the linear fits to the correlation
with a theoretical constraint by the deuteron data. The qualities of the fits are good,
with small $\chi^2/ndf=0.897$ and $\chi^2/ndf=1.18$ for the fit to the data
from J. Arrington et al. and the fit to the data from S. Malace et al., respectively.
The slopes of the fitted linear functions are obtained to be $0.039\pm0.002$ and $0.041\pm0.002$
for Fit 3 and 4, respectively (see Fig. \ref{fig1}(b)). Therefore the formula for the correlation
between the EMC slope and RSIE per nucleon is written as
\begin{equation}
\left| \frac{dR_{EMC}}{dx} \right|=(\frac{RSIE}{A~\mbox{MeV}}-1.112)\times (0.041\pm0.002).
\label{EMC-RSIE}
\end{equation}
The slope of the linear correlation in above formula is obtained from Fit 4.
Fit 4 is the linear fit to the global analysis data of EMC slopes from many experimental
measurements \cite{review4}. The correlation between RSIE and the EMC slope hints that
nuclear force plays an important role in the EMC effect.

With the assumption that EMC slope and RSIE are linearly correlated, the amazing correlation
allows us to extract significant information about the deuteron. In the EMC effect
measurements, the deuteron is used as the denominator, which is often viewed as
an good approximation to a free proton and neutron system. In recent state-of-the-art
measurement \cite{BONuS}, model-independent structure function of free neutron is extracted
from the deuteron target being not treated as a free proton and neutron.
Although the deuteron is loosely bound, its structure function is different from
that of a free proton and neutron system. The in-medium correction (IMC) effect,
defined as $\frac{\sigma_{A}/A}{(\sigma_p+\sigma_n)/2}$, was extracted for the deuteron
using the correlation between the EMC effect and SRC \cite{detail-emc-src,src-emc1,src-emc2}.
Similarly, the deuteron IMC effect can be extracted from the extrapolation of the EMC effect
to the free $pn$ pair where the residual strong interaction energy is zero.
The intercept is extracted to be $-0.046\pm 0.003$ from the linear fit to the correlation
between the EMC effect and RSIE per nucleon. Thus the IMC slope of the deuteron is given
\begin{equation}
\left| \frac{dR_{IMC}(d)}{dx} \right|= 0.046\pm0.003.
\label{IMC-d}
\end{equation}
The obtained deuteron IMC slope is very close to the fitted value of the local
density explanation. The IMC slope for the deuteron was yielded to be $0.051\pm0.003$
by the local density fit \cite{detail-emc-src}. The EMC-RSIE correlation is
consistent with the local density assumption in terms of the IMC effect for the deuteron.

This correlation also allows us to predict the size of the EMC effect for unmeasured
nuclei by simply using Eq. (\ref{EMC-RSIE}), owning to the comprehensively and precisely
measured nuclear binding energy data. Further EMC measurements on very light nuclei
would be useful to test EMC-RSIE correlation. The approved JLab E12-10-008 \cite{E1210008}
and E12-10-103 \cite{E1210103} experiments for the 12 GeV physics program will measure
the nuclear EMC effect from very light to medium heavy nuclei. By applying EMC-RSIE
correlation, the predicted EMC slopes of the nuclei which will be measured at JLab are
shown in Table \ref{table2}. A comparison of the EMC effect on $^{40}$Ca and $^{48}$Ca
will be made in E12-10-008 experiment for the first time. From the EMC-RSIE correlation,
the EMC slope of $^{48}$Ca is slightly smaller than that of $^{40}$Ca, though atomic mass
number of $^{48}$Ca is larger. The EMC effect for $^3$H and $^3$He mirror nuclei will be
measured in E12-10-103 experiment. The predicted EMC slope of tritium is very close to
that of $^3$He. JLab proposed experiments provide a good opportunity to test the correlation
between the EMC effect and RSIE.

\begin{table}[htp]
\begin{center}
\caption{
The predicted EMC slopes of various nuclei which will be measured in
JLab E12-10-008 \cite{E1210008} and E12-10-103 \cite{E1210103} experiments.
The estimated errors are from the uncertainty of the linear fit to the EMC-RSIE correlation.
}
\begin{tabular}{cccc}
\hline
Nucleus & $|dR_{EMC}/dx|$ & Nucleus &  $|dR_{EMC}/dx|$  \\
\hline
$^3$H   & 0.070$\pm$0.004 & $^{10}$B & 0.247$\pm$ 0.012  \\
$^3$He  & 0.073$\pm$0.004 & $^{11}$B & 0.262$\pm$ 0.013   \\
$^4$He  & 0.254$\pm$0.013 & $^{12}$C  & 0.301$\pm$0.015   \\
$^6$Li  & 0.189$\pm$0.010 & $^{40}$Ca & 0.386$\pm$0.019    \\
$^7$Li  & 0.197$\pm$0.010 & $^{48}$Ca & 0.373$\pm$0.019  \\
$^9$Be  & 0.238$\pm$0.012 & $^{63}$Cu & 0.408$\pm$0.020  \\
\hline
\end{tabular}
\label{table2}
\end{center}
\end{table}

\section{Discussions}
\label{sec4}

\begin{figure}[htp]
\centering
\includegraphics[width=0.46\textwidth]{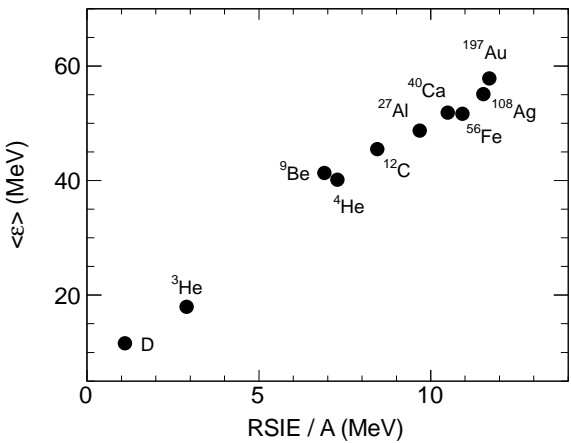}
\caption{
Correlation between the average nucleon separation energy \cite{Kulagin-data}
and RSIE per nucleon.
}
\label{fig2}
\end{figure}

The off-shell correction to the bound nucleon quantified by the nucleon separation energy
describes well the EMC effect of both heavy and light nuclei. Hence it is interesting
to look for the correlation between the average nucleon separation energy and RSIE.
The average nucleon separation energy \cite{Kulagin-data} as a function of the per-nucleon
RSIE is shown in Fig. \ref{fig2}. Roughly, RSIE per nucleon and the average nucleon
separation energy are linearly correlated. In fact, these two quantities are both related
to the nuclear force. Compared to the average separation energy, the RSIE describes
only the strong interaction part of the nuclear force. In addition, RSIE is easy to calculate.

\begin{figure}[htp]
\centering
\includegraphics[width=0.46\textwidth]{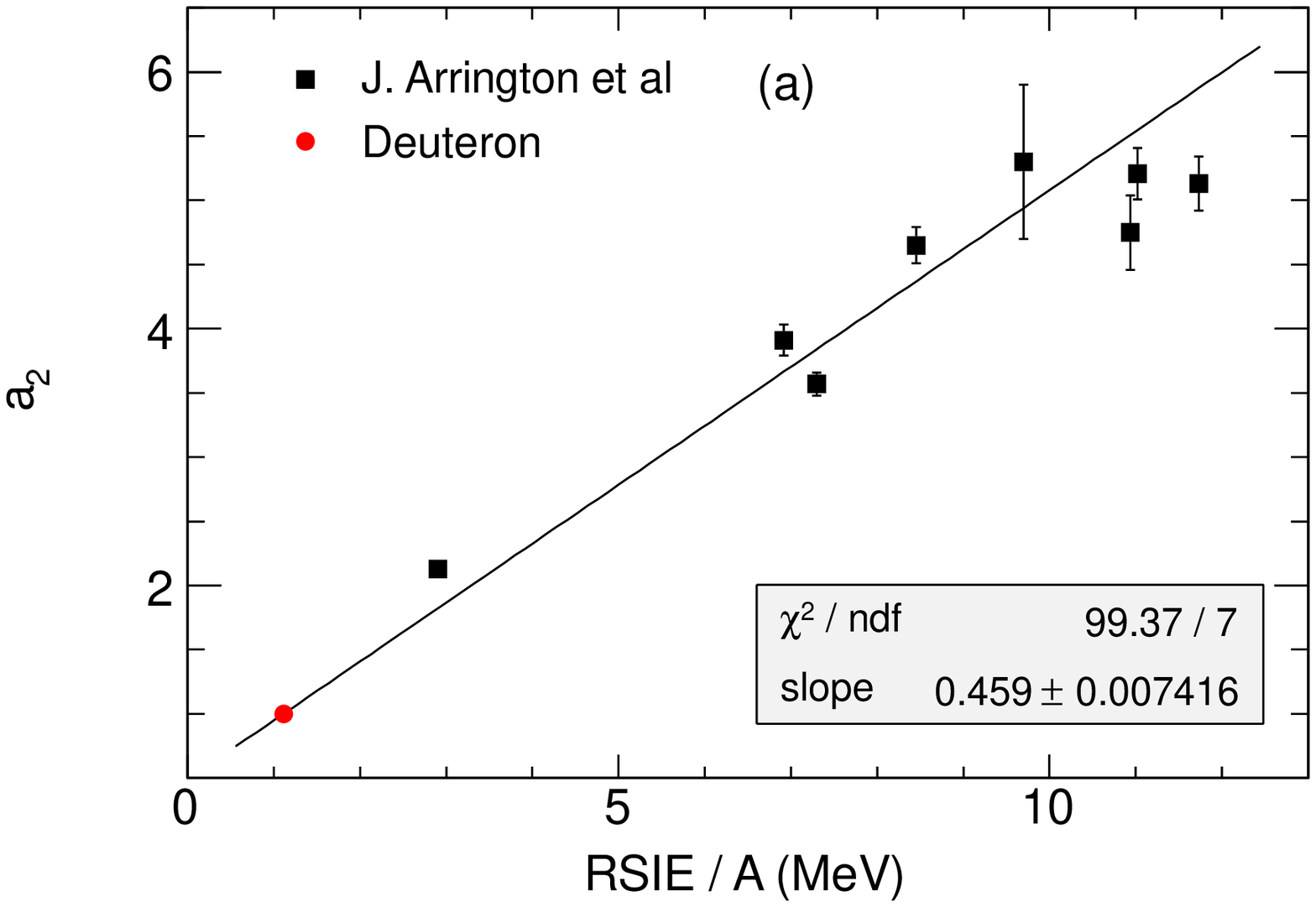}
\includegraphics[width=0.46\textwidth]{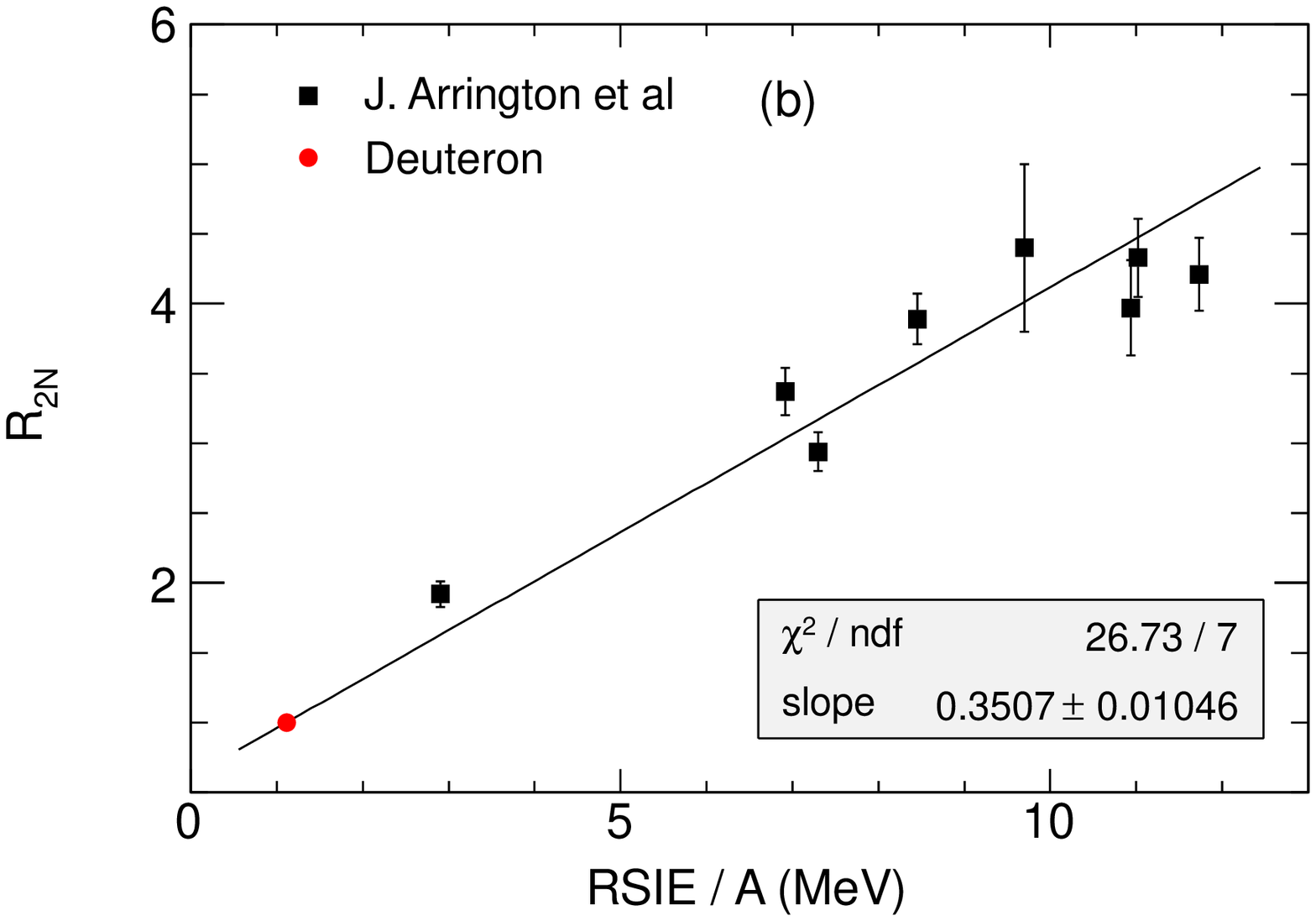}
\includegraphics[width=0.46\textwidth]{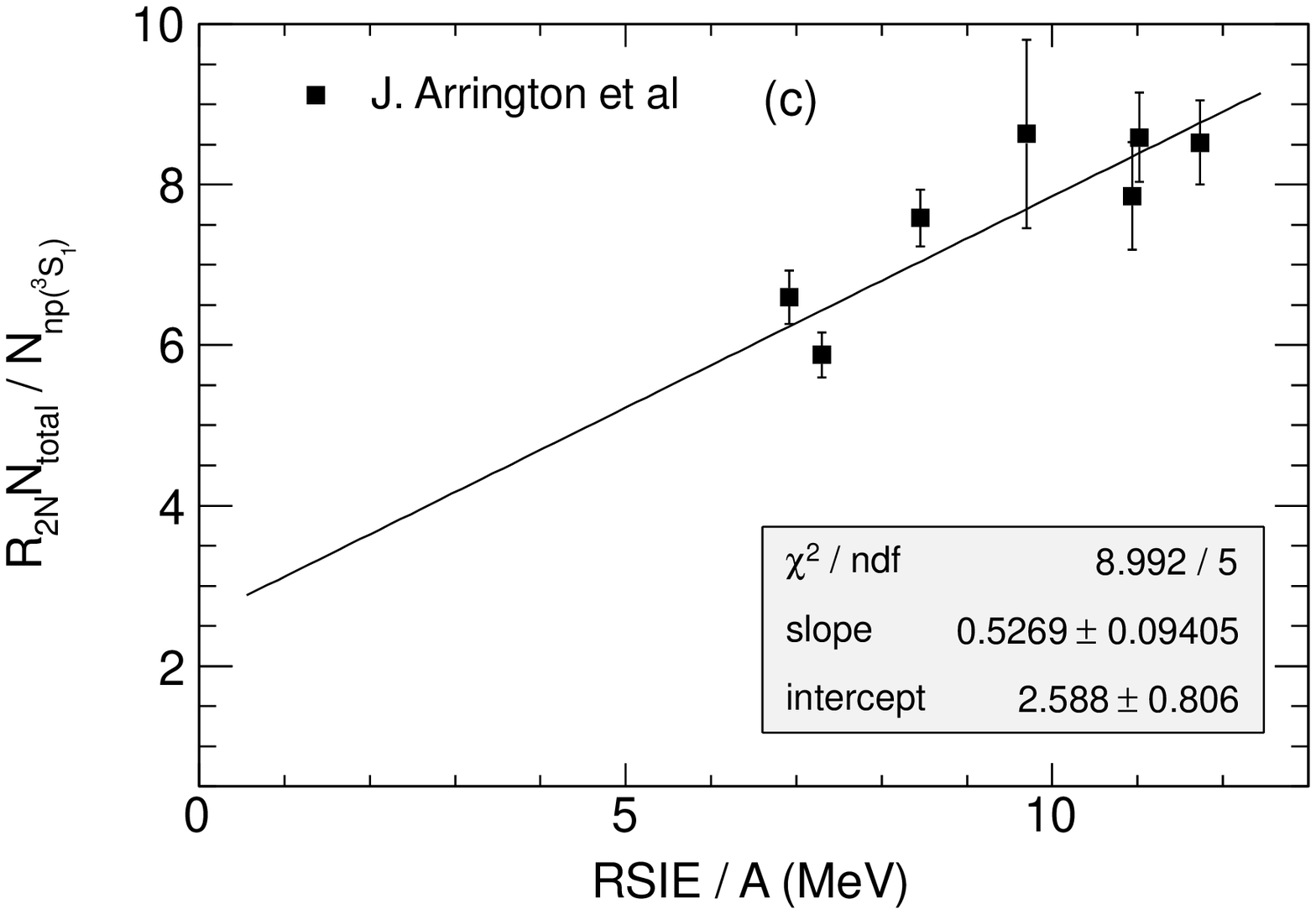}
\caption{
(Color online.)
Plot (a) shows $a_2$ versus per-nucleon RSIE;
Plot (b) shows $R_{2N}$ versus per-nucleon RSIE;
Plot (c) shows $R_{2N}N_{total}/N_{np(^3S_1)}$ versus per-nucleon RSIE.
}
\label{fig3}
\end{figure}

It is reasonable to investigate the relationship between RSIE and nucleon-nucleon (NN)
SRC, as there is strong connection found between SRC and the EMC effect.
The strength of NN SRC is usually described by $a_2$ or $R_{2N}$.
We take the combined data of $a_2$ and $R_{2N}$ from Ref. \cite{detail-emc-src}
in the analysis. Fig. \ref{fig3}(a) shows NN SRC scaling factor $a_2$ as a function
of RSIE per nucleon. There is no clear linear correlation observed for these two quantities.
A linear fit with the theoretical constraint by the deuteron as well as the quality
of the fit ($\chi^2/ndf$) are shown in the figure. Fig. \ref{fig3}(b) shows SRC ratio
$R_{2N}$ as a function of RSIE per nucleon. No obvious linear correlation is shown,
however the $\chi^2/ndf=3.82$ becomes smaller. If $nn$, $np$, and $pp$ pairs all have
equal probability to form high local density configurations, $R_{2N}N_{total}/N_{np(^3S_1)}$
($N_{total}=N_{nn}+N_{np}+N_{pp}$) should expected to be a better description for
the local nuclear environment, as the $A(e,e')$ experiment used to extract $R_{2N}$
are mostly sensitive to the tensor part of the correlations and those only couple to
spin 1 pairs ($np(^3S_1)$) \cite{detail-emc-src,Maarten11,Maarten12}.
We take the number of $nn$, $pp$, $np(^1S_0)$ and $np(^3S_1)$ pairs from
Refs. \cite{Maarten11,Maarten12,Wim}. The correlation between $R_{2N}N_{total}/N_{np(^3S_1)}$
and per-nucleon RSIE is shown in Fig. \ref{fig3}(c). Basically, the per-nucleon RSIE
and $R_{2N}N_{total}/N_{np(^3S_1)}$ are linearly correlated. The line in Fig. \ref{fig3}(c)
is an linear fit to the correlation. The data of the deuteron is not included in the fit,
for we do not know $R_{2N}N_{total}/N_{np(^3S_1)}$ of the deuteron.
The method \cite{Maarten11,Maarten12} of calculating number of correlated nucleon pairs
is not accurate for two body systems.

\begin{figure}[htp]
\centering
\includegraphics[width=0.46\textwidth]{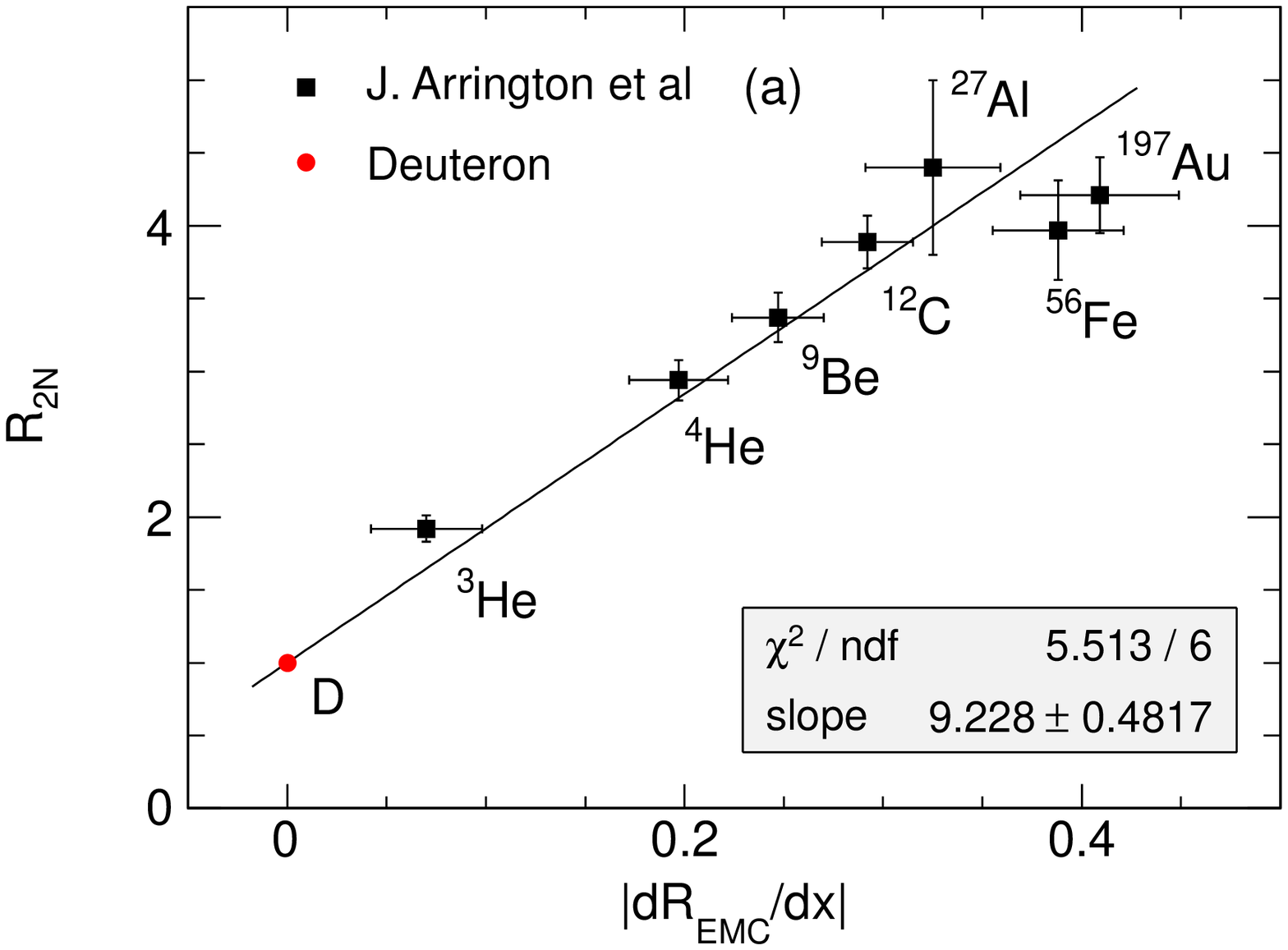}
\includegraphics[width=0.46\textwidth]{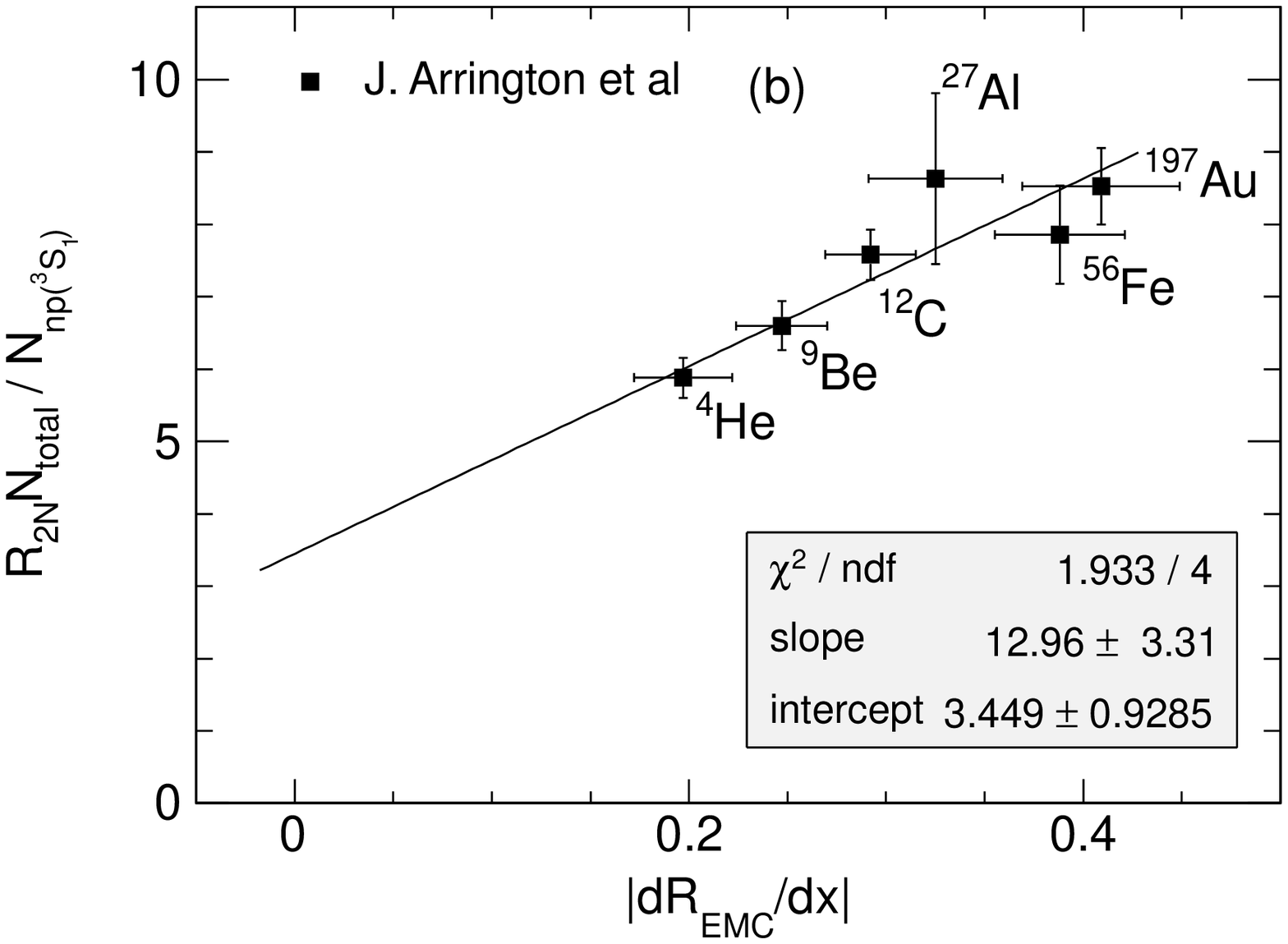}
\caption{
(Color online.)
Plot (a) shows $R_{2N}$ versus EMC slopes;
Plot (b) shows $R_{2N}N_{total}/N_{np(^3S_1)}$ versus EMC slopes.
}
\label{fig4}
\end{figure}

If $R_{2N}$ or $R_{2N}N_{total}/N_{np(^3S_1)}$ scales with per-nucleon RSIE,
we should also find the correlation between $R_{2N}$ and $|dR_{EMC}/dx|$
or the correlation between $R_{2N}N_{total}/N_{np(^3S_1)}$ and $|dR_{EMC}/dx|$.
The correlations between $R_{2N}$ and EMC slope, and between $R_{2N}N_{total}/N_{np(^3S_1)}$
and EMC slope are shown in Fig. \ref{fig4}. Amazingly, linear fits to the correlations
both show good quality of fit. One simple explanation for the correlations among RSIE,
EMC slope and $R_{2N}N_{total}/N_{np(^3S_1)}$ is as follows. The narrower the
repulsive core of nuclear force is, the larger RSIE is and the stronger SRC ratio is.
Narrower repulsive core leads to higher local nuclear density so as to
enhance the strength of the EMC effect. Anyhow, the analysis shows correlations
among RSIE, EMC slope, and $R_{2N}N_{total}/N_{np(^3S_1)}$.

\section{Summary}
\label{sec5}

It is shown that the magnitude of the EMC effect is linearly correlated to RSIE per nucleon.
The quality of the fit to the correlation between the EMC effect and RSIE is close to
the quality of the fit to the correlation between the EMC effect and SRC.
Note that the error of RSIE is not included in the fit while errors of both $a_2$ and $R_{2N}$
are included in the linear fits. The IMC slope of deuteron is extracted
to be $0.046\pm 0.002$, which is consistent with the value obtained from local
density explanation fit. Assuming that the EMC effect is linearly correlated with RSIE per nucleon,
we predicted the EMC slopes $|dR_{EMC}/dx|$ of various nuclei which will be measured at JLab.
Comparing the correlation between EMC slope and nuclear binding energy with
the correlation between EMC slope and RSIE, we find that it is the strong interaction
part of the binding that the magnitude of the EMC effect scales with.
The EMC effect is a QCD effect rather than the binding effect. The connection between RSIE
and the EMC effect is crucial for unveiling the underlying mechanism of the EMC effect.

Various correlations related to the EMC effect are observed. We should be careful
when we try to explain these correlations. While it an open question to understand
the RSIE dependence of the EMC effect, nuclear force surely plays an important role
in the EMC effect. In the picture of nucleon-nucleon potential, the potential minimum
is at distances of less than 0.9 femtometer from both phenomenological potential
\cite{Reid93,AV18,Born} and Lattice QCD calculations \cite{LQCD1,LQCD3}.
No doubt that nuclear force is of short range and RSIE is sensitive to average
local nuclear density. The observed correlation between RSIE per nucleon and the EMC slope
supports that the local nuclear environment is an important factor for the EMC effect.

\section*{Acknowledgments}
We are grateful for the fruitful discussions with W. Zhu, the average nucleon separation
energy data from S. A. Kulagin, and the data of the number of SRC pairs from W. Cosyn.
This work was supported by the National Basic Research Program of China (973 Program)
2014CB845406, the National Natural Science Foundation of China under
Grant Number 11175220 and Century Program of Chinese Academy of Sciences Y101020BR0.

\end{document}